\documentclass[prd,superscriptaddress,amsfonts,amssymb,amsmath,showpacs,twocolumn]{revtex4-2}
\usepackage{bm}
\usepackage{amsfonts}
\usepackage{latexsym}
\usepackage[latin1]{inputenc}
\usepackage{graphicx}
\usepackage{amsmath}
\usepackage{palatino}
\usepackage{mathpazo}
\usepackage{textcomp}
\usepackage{float}
\usepackage{booktabs}
\usepackage{dcolumn}
\usepackage{hyperref}
\hypersetup{colorlinks,citecolor=blue}
\usepackage{amsmath}
\usepackage{xcolor}
\usepackage{orcidlink}
\usepackage[caption=false]{subfig}
\usepackage{commath}
\def\jnl@style{\it}
\def\aaref@jnl#1{{\jnl@style#1}}

\def\aaref@jnl#1{{\jnl@style#1}}

\def\aj{\aaref@jnl{AJ}}                   % Astronomical Journal
\def\apj{\aaref@jnl{ApJ}}                 % Astrophysical Journal
\def\apjl{\aaref@jnl{ApJ}}                % Astrophysical Journal, Letters
\def\apjs{\aaref@jnl{ApJS}}               % Astrophysical Journal, Supplement
\def\apss{\aaref@jnl{Ap\&SS}}             % Astrophysics and Space Science
\def\aap{\aaref@jnl{A\&A}}                % Astronomy and Astrophysics
\def\aapr{\aaref@jnl{A\&A~Rev.}}          % Astronomy and Astrophysics Reviews
\def\aaps{\aaref@jnl{A\&AS}}              % Astronomy and Astrophysics, Supplement
\def\mnras{\aaref@jnl{Mon.~Not.~Roy.~Astron.~Soc.}}             % Monthly Notices of the RAS
\def\prd{\aaref@jnl{Phys.~Rev.~D}}        % Physical Review D
\def\prc{\aaref@jnl{Phys.~Rev.~C}}  % Physical Review C
\def\prl{\aaref@jnl{Phys.~Rev.~Lett.}}    % Physical Review Letters
\def\qjras{\aaref@jnl{QJRAS}}             % Quarterly Journal of the RAS
\def\skytel{\aaref@jnl{S\&T}}             % Sky and Telescope
\def\ssr{\aaref@jnl{Space~Sci.~Rev.}}     % Space Science Reviews
\def\zap{\aaref@jnl{ZAp}}                 % Zeitschrift fuer Astrophysik
\def\nat{\aaref@jnl{Nature}}              % Nature
\def\aplett{\aaref@jnl{Astrophys.~Lett.}} % Astrophysics Letters
\def\apspr{\aaref@jnl{Astrophys.~Space~Phys.~Res.}} % Astrophysics Space Physics Research
\def\physrep{\aaref@jnl{Phys.~Rep.}}      % Physics Reports
\def\physscr{\aaref@jnl{Phys.~Scr}}       % Physica Scripta
\def\commat{\aaref@jnl{Comm.~Math.~Phys.}}              % Communications in Mathematical Physics
\def\science{\aaref@jnl{Science}}               % Science
\def\cqg{\aaref@jnl{Classical Quant.~Grav.}}            % Classical and Quantum Gravity
\def\jpcs{\aaref@jnl{JPCS}}                                     % Journal of Physics Conference Series
\def\ijmpd{\aaref@jnl{Int.~J.~Mod.~Phys.~D}}                    % International Journal of Modern Physics D
\def\grg{\aaref@jnl{Gen.~Relat.~Gravit.}}               % General Relativity and Gravitation
\def\rpp{\aaref@jnl{Rep.~Prog.~Phys.}}          % Reports on Progress in Physics
\def\npa{\aaref@jnl{Nucl.~Phys.~A}}        % Nuclear Physics A
\def\lrr{\aaref@jnl{Living Rev.~Rel.}}                   % Living reviews in relativity
\def\jcap{\aaref@jnl{J.~Cosmology Astropart.~Phys.}}    % Journal of cosmology and astroparticle physics
\def\rmp{\aaref@jnl{Rev.~Mod.~Phys.}}   %Reviews of modern physics
\def\epjc{\aaref@jnl{Eur.~Phys.~J.~C}}

\allowdisplaybreaks[1]
\begin{document}

\color{black}       %% For one column

\title{Anisotropic nature of space-time in $f\left( Q\right) $ gravity}
%\end{document}
\author{M. Koussour\orcidlink{0000-0002-4188-0572}}
\email{pr.mouhssine@gmail.com}
\affiliation{Quantum Physics and Magnetism Team, LPMC, Faculty of Science Ben
M'sik,\\
Casablanca Hassan II University,
Morocco.}

\author{S.H. Shekh\orcidlink{0000-0003-4545-1975}}
\email{da\_salim@rediff.com}
\affiliation{Department of Mathematics. S. P. M. Science and Gilani Arts Commerce
College,\\ Ghatanji, Dist. Yavatmal, Maharashtra-445301, India.}

\author{M. Bennai\orcidlink{0000-0003-1424-7699}}
\email{mdbennai@yahoo.fr }
\affiliation{Quantum Physics and Magnetism Team, LPMC, Faculty of Science Ben
M'sik,\\
Casablanca Hassan II University,
Morocco.} 
\affiliation{Lab of High Energy Physics, Modeling and Simulations, Faculty of
Science,\\
University Mohammed V-Agdal, Rabat, Morocco.}
%
%%%%%%%%%%%%%%%%%%%%%%%%%%%%%%%%%%%%%  DATE  %%%%%%%%%%%%%%%%%%%%%%%%%%%%%%%%%%%%
\date{\today}
\begin{abstract}
In this paper, we consider the Bianchi type-I space-time with perfect fluid
as matter content of the Universe in the framework of $f(Q)$ gravity (where $%
Q$ is the non-metricity scalar) recently proposed by Jim\'{e}nez et al.
(Phys. Rev. D 98.4 (2018): 044048). We find exact solutions of the field
equations using the anisotropic property of space-time for volumetric hybrid
expansion. The cosmological parameters for the linear form of non-metricity
scalar i.e. $f\left( Q\right) =\alpha Q+\beta $ (where $\alpha $ and $\beta $
are free model parameters) are discussed and compared with recent Hubble
measurements. Also, we have obtained the best fitting values of the model
parameters $k, n$ and $H_{0}$ by constraining our model with updated Hubble
datasets consisting of 57 data points ($31$ (DA) and $26$ (BAO+other)) along
with also examined the stability of the model and test its validity by
energy conditions.\newline
\end{abstract}

\maketitle

\date{\today}
\section{Introducion}
\label{sec1}

The theory of general relativity (GR) is the most famous theory of gravity
published by Albert Einstein in 1915 \cite{ref1}, for having been
experimentally tested more than once, and all the results are surprisingly
consistent with the theory's predictions \cite{ref2, ref3, ref4}. According
to this theory, gravity is just a curvature of space-time due to the
presence of a large mass, and therefore, this theory linked the matter
content of the Universe with the geometry of the fabric of the space-time.
The geometry of space-time is represented by the Einstein tensor $G_{\mu \nu
}$, and matter is represented by the energy-momentum tensor $T_{\mu \nu }$,
and thus we get the Einstein field equations as $G_{\mu \nu }=kT_{\mu \nu }$
(in natural units $k=8\pi Gc^{-4}=1$). These equations can be derived by
means of a mathematical object called Einstein-Hilbert (EH) action which is
given as follows

\begin{equation}
S_{EH}=\int \left[ \frac{1}{2}R+\mathcal{L}_{m}\right] d^{4}x\sqrt{-g},
\label{eqn1}
\end{equation}%
where $g=\det \left( g_{\mu \nu }\right) $ is the determinant of the metric
tensor $g_{\mu \nu }$, $R$ is the Ricci scalar and $\mathcal{L}_{m}$ is the
matter Lagrangian density. GR is based on an important property that the
covariant derivative $\nabla _{\gamma }$ of the metric is zero $\left(
\nabla _{\gamma }g_{\mu \nu }=0\right) $ \cite{ref5}. Despite all the
success of this theory, there are major common questions, both theoretical
and experimental, for which GR has not found an explanation, such as the
initial singularity, dark matter, dark energy (DE), etc. In this paper, we
will discuss the DE problem. The DE is an exotic form of energy with
negative pressure or negative equation of state (EoS) parameter, because $%
\omega =\frac{p}{\rho }$, where $\rho $ represents the energy density of the
Universe and $p$ represents pressure, and it behaves like a repulsive
gravity. The presence of DE in the Universe is supported by two of most
important modern cosmological observational facts. The first is the
composition of our Universe, this shows that dark energy contributes $68.3\%$
of the total energy density (the rest is dark matter and baryonic matter) 
\cite{ref6}. The second is strong evidence coming mainly from type-Ia
supernovae (SN-Ia) observations that the Universe is today undergoing an
accelerated expansion phase \cite{ref7, ref8}. In order to obtain the
accelerating expansion, the second derivative of the scale factor must be
positive (i.e. $\overset{..}{a}>0$) or in an equivalent way the deceleration
parameter is negative ( i.e. $q=-\frac{a\overset{..}{a}}{\overset{.}{a}^{2}}%
<0$). The simplest candidate for DE is the cosmological constant $\Lambda $,
which is regarded to be equivalent with the vacuum energy. But soon other
problems appeared, such as the coincidence and magnitude \cite{ref9}.
Therefore, many researchers have searched for other interpretations about
the origin of DE. In fact, two distinct approaches have been proposed for
treating the DE problem. The first approach deals with GR where an
additional energy component is introduced to explain the cosmic acceleration
phenomenon. In this context, several alternative DE models have been
suggested, but so far no suitable candidate has been found. The other
approach is to modify the EH action. In this context, the main idea was to
replace the Ricci scalar by a function of $R$, which is known as $f\left(
R\right) $ gravity \cite{ref10, ref11}. Other modified theories have been
proposed like $f\left( T\right) $ gravity (where $T$ is the torsion), $%
f\left( G\right) $ gravity (where $G$ is the Gauss-Bonnet),\ see these works
in this regard \cite{ref12, ref13, ref14}.

Recently, Jim\'{e}nez et al. \cite{ref15}\ proposed a new modified theory of
gravity called $f\left( Q\right) $\ gravity where $Q$ is the non-metricity
scalar which attracted the attention of many researchers. This theory is
based on Weyl geometry, which is a generalization of Riemannian geometry
that is the mathematical basis of GR. In Weyl geometry, gravitational
effects do not occur because of the change in direction of a vector in
parallel transport, but because of the change in length of the vector itself 
\cite{ref16}. Generally, we can categorize gravitational interactions in the
space-time variety into three types of geometrical objects, specifically the
curvature, torsion, and non-metricity. In GR, gravitational interactions are
caused by the curvature of space-time. Accordingly, $f\left( R\right) $
gravity is a modification of GR based on curvature with zero torsion and
non-metricity. Likewise, $f\left( T\right) $ gravity is a modification of
torsion-based gravity (the teleparallel equivalent of GR) with zero
curvature and non-metricity. Finally, $f\left( Q\right) $\ gravity is a
modification of the symmetric teleparallel equivalent of GR with zero
curvature and torsion \cite{ref17, ref18}. In differential geometry and to
be more precise in Weyl-Cartan geometry, we need the symmetric metric tensor 
$g_{\mu \nu }$ to define the length of a vector, and an asymmetric
connection $\widetilde{{\Gamma }}{^{\gamma }}_{\mu \nu }$\ in order to
defines the covariant derivatives and parallel transport. Thus, the general
affine connection can be decomposed into three parts: the Christoffel symbol 
${\Gamma ^{\gamma }}_{\mu \nu }$, the contortion tensor ${C^{\gamma }}_{\mu
\nu }$, and the disformation tensor ${L^{\gamma }}_{\mu \nu }$,
respectively, which are written as \cite{ref16}

\begin{equation}
\widetilde{{\Gamma }}{^{\gamma }}_{\mu \nu }={\Gamma ^{\gamma }}_{\mu \nu }+{%
C^{\gamma }}_{\mu \nu }+{L^{\gamma }}_{\mu \nu },  \label{eqn2}
\end{equation}%
where the Levi-Civita connection of the metric $g_{\mu \nu }$ is given by

\begin{equation}
{\Gamma ^{\gamma }}_{\mu \nu }\equiv \frac{1}{2}g^{\gamma \sigma }\left( 
\frac{\partial g_{\sigma \nu }}{\partial x^{\mu }}+\frac{\partial g_{\sigma
\mu }}{\partial x^{\nu }}-\frac{\partial g_{\mu \nu }}{\partial x^{\sigma }}%
\right) ,  \label{eqn3}
\end{equation}%
the contorsion tensor ${C^{\gamma }}_{\mu \nu }$ is

\begin{equation}
{C^{\gamma }}_{\mu \nu }\equiv \frac{1}{2}{T^{\gamma }}_{\mu \nu }+T_{(\mu
}{}^{\gamma }{}_{\nu )},  \label{eqn4}
\end{equation}%
with ${T^{\gamma }}_{\mu \nu }\equiv 2{\Gamma ^{\gamma }}_{[\mu \nu ]}$ in
the above equation is the torsion tensor. Lastly, the disformation tensor ${%
L^{\gamma }}_{\mu \nu }$ is obtained from the non-metricity tensor as

\begin{equation}
{L^{\gamma }}_{\mu \nu }\equiv \frac{1}{2}g^{\gamma \sigma }\left( Q_{\nu
\mu \sigma }+Q_{\mu \nu \sigma }-Q_{\gamma \mu \nu }\right) .  \label{eqn5}
\end{equation}

In Eq. (\ref{eqn5}), the non-metricity tensor $Q_{\gamma \mu \nu }$ is
defined as the (minus) covariant derivative of the metric tensor with regard
to the Weyl-Cartan connection $\widetilde{\Gamma }_{\mu \nu }^{\lambda }$,
i.e. $Q_{\gamma \mu \nu }=\nabla _{\gamma }g_{\mu \nu }$, and it can be
gained

\begin{equation}
Q_{\gamma \mu \nu }=-\frac{\partial g_{\mu \nu }}{\partial x^{\gamma }}%
+g_{\nu \sigma }\widetilde{{\Gamma }}{^{\sigma }}_{\mu \gamma }+g_{\sigma
\mu }\widetilde{{\Gamma }}{^{\sigma }}_{\nu \gamma }.  \label{eqn6}
\end{equation}

In the case of a flat and torsion-free connection, the connection (\ref{eqn2}%
) can be parameterized as

\begin{equation}
\widetilde{{\Gamma }}{^{\gamma }}_{\mu \beta }=\frac{\partial x^{\gamma }}{%
\partial \xi ^{\rho }}\partial _{\mu }\partial _{\beta }\xi ^{\rho }.
\label{eqn7}
\end{equation}

In Eq. (\ref{eqn7}), $\xi ^{\gamma }=$ $\xi ^{\gamma }\left( x^{\mu }\right) 
$ is an invertible relation. Therefore, it is constantly potential to get a
coordinate system so that the connection $\widetilde{\Gamma }_{\mu \nu
}^{\gamma }$ disappears. This condition is called coincident gauge and the
covariant $\nabla _{\gamma }$\ derivative reduces to the partial derivative $%
\partial _{\gamma }$ \cite{ref18}. Hence, in the coincident gauge
coordinate, we get

\begin{equation}
Q_{\gamma \mu \nu }=\partial _{\gamma }g_{\mu \nu }.  \label{eqn8}
\end{equation}

The symmetric teleparallel gravity is a geometric description of gravity
equivalent to GR under coincident gauge coordinates in which $\widetilde{{%
\Gamma }}{^{\gamma }}_{\mu \nu }=0$ and ${C^{\gamma }}_{\mu \nu }=0$, and
therefore from Eq. (\ref{eqn2}) we can conclude that ${\Gamma ^{\gamma }}%
_{\mu \nu }=-{L^{\gamma }}_{\mu \nu }$ \cite{ref16}. This theory has
received great interest from many authors because it gives good results for
several issues. The first cosmological solutions in $f(Q)$ gravity emerge in
Ref. \cite{ref19}. Next, Mandal et al. presented a complete test of energy
conditions for $f\left( Q\right) $ gravity models and constraint families of 
$f\left( Q\right) $ models compatible with the current accelerating
expansion of the Universe \cite{ref20}. The cosmography in $f\left( Q\right) 
$ gravity is also discussed by Mandal et al. in another work \cite{ref21}.
The growth index of matter perturbations has been examined in the context of 
$f\left( Q\right) $ gravity in \cite{ref22}. The geodesic deviation equation
in $f\left( Q\right) $ gravity has been explored and some fundamental
results were gained in \cite{ref23}. Harko et al. discussed the coupling
matter in modified $f(Q)$ gravity \cite{ref24}. Dimakis et al. investigated
quantum cosmology for a $f\left( Q\right) $ polynomial model \cite{ref25}.
Holographic dark energy models in $f\left( Q\right) $ gravity are discussed
by Shekh \cite{ref26}.~To study the early evolution of the Universe, A. De
et al. obtained the Bianchi-I-type field equations in $f\left( Q\right) $
gravity and some cosmological parameters are discussed in this context \cite%
{ref27} and, several other issues in $f\left( Q\right) $ gravity are
discussed in Refs. \cite{ref28, ref29, ref30, ref31}. Motivated from the
studies outlined above, in this paper, we have investigated the perfect
fluid solutions for Bianchi type-I space-time in $f\left( Q\right) $
gravity. Also, the modified Friedmann equations are obtained by applying the
anisotropic nature of space-time. Finally, we have discussed some of the
physical parameters of the model and have compared with recent Hubble
measurements.

The current article is organized as follows: In Sec. \ref{sec2}, we present the
basic formalism of $f\left( Q\right) $ gravity and field equations of the
Bianchi type-I space-time. We establish the solutions of the anisotropic
cosmological model with the choice of a linear $f\left( Q\right) $ function
i.e. $f\left( Q\right) =\alpha Q+\beta$, and examined some physical
parameters in Sec. \ref{sec3}. In Sec. \ref{sec4}, we consider the stability analysis of
the anisotropic model by means of the squared sound speed, and we test the
validity of the model using energy conditions. We give some physical
parameters in terms of redshift in Sec. \ref{sec5}. Finally, we discuss the best fit
values of model parameters from observation in Sec. \ref{sec6} and in sec. \ref{sec7} we
presents the conclusion and discussion.\newline

\section{Field equations in $f\left( Q\right) $ gravity}
\label{sec2}

In modified symmetric teleparallel gravity or $f\left( Q\right) $ gravity,\
the action is given by \cite{ref19}

\begin{equation}
S=\int \left[ \frac{1}{2}f(Q)+\mathcal{L}_{m}\right] d^{4}x\sqrt{-g},
\label{eqn9}
\end{equation}%
where $f(Q)$ is an arbitrary function of the non-metricity scalar $Q$. The
non-metricity tensor and its traces are obtained as

\begin{equation}
Q_{\gamma \mu \nu }=\nabla _{\gamma }g_{\mu \nu }\,,  \label{eqn10}
\end{equation}%
\begin{equation}
Q_{\gamma }={{Q_{\gamma }}^{\mu }}_{\mu }\,,\qquad \widetilde{Q}_{\gamma }={%
Q^{\mu }}_{\gamma \mu }\,.  \label{eqn11}
\end{equation}

Moreover, the superpotential tensor (non-metricity conjugate) is given by 
\begin{equation}
4{P^{\gamma }}_{\mu \nu }=-{Q^{\gamma }}_{\mu \nu }+2Q_{({\mu ^{^{\gamma }}}{%
\nu })}-Q^{\gamma }g_{\mu \nu }-\widetilde{Q}^{\gamma }g_{\mu \nu }-\delta _{%
{(\gamma ^{^{Q}}}\nu )}^{\gamma }\,,  \label{eqn12}
\end{equation}

Next, the trace of the non-metricity tensor can be obtained as 
\begin{equation}
Q=-Q_{\gamma \mu \nu }P^{\gamma \mu \nu }\,.  \label{eqn13}
\end{equation}

Furthermore, the matter energy-momentum tensor is of the form

\begin{equation}
T_{\mu \nu }=-\frac{2}{\sqrt{-g}}\frac{\delta (\sqrt{-g}\mathcal{L}_{m})}{%
\delta g^{\mu \nu }}\,.  \label{eqn14}
\end{equation}

The field equations in symmetric teleparallel gravity corresponding to
action (\ref{eqn9}) are 
\begin{widetext}
\begin{equation}
\frac{2}{\sqrt{-g}}\nabla _{\gamma }\left( \sqrt{-g}f_{Q}P^{\gamma }{}_{\mu
\nu }\right) +\frac{1}{2}fg_{\mu \nu }+f_{Q}\left( P_{\mu \gamma i}Q_{\nu
}{}^{\gamma i}-2Q_{\gamma i\mu }P^{\gamma i}{}_{\nu }\right) =-T_{\mu \nu },
\label{eqn15}
\end{equation}%
\end{widetext} where the subscript marks the derivative with respect to $Q$.
In addition, we can also take the variation of (\ref{eqn9}) with respect to
the connection, which gives

\begin{equation}
\nabla _{\mu }\nabla _{\gamma }\left( \sqrt{-g}f_{Q}P^{\gamma }{}_{\mu \nu
}\right) =0.  \label{eqn16}
\end{equation}

We consider the matter content of the Universe as a perfect fluid which
energy-momentum tensor is

\begin{equation}
T_{\mu \nu }=\left( \rho +p\right) u_{\mu }u_{\nu }+pg_{\mu \nu },
\label{eqn17}
\end{equation}%
where $\rho $ and$\ p$ are the energy density and pressure of the matter
content. The four-velocity vector $u^{\mu }$ is presumed to satisfy $u^{\mu
}u_{\mu }=-1$.

The Bianchi type-I space-time is a single of the simplest form of
anisotropic space-time, which represents a homogeneous and spatially flat
space-time. Hence, In this paper, we consider Bianchi type-I space-time
which is the direct generalization of the flat FLRW space-time and has the
form 
\begin{equation*}
ds^{2}=-dt^{2}+A^{2}(t)dx^{2}+B^{2}(t)\left( dy^{2}+dz^{2}\right) ,
\end{equation*}%
where metric potentials $A\left( t\right) $ and $B\left( t\right) $ depend
only on cosmic time $\left( t\right) $ and flat FLRW space-time can be
attained if we set $A\left( t\right) =B\left( t\right) =a\left( t\right) $.
For perfect fluid as matter contents, the corresponding field equations for
Bianchi type-I space-time are derived as \cite{ref27} 
\begin{widetext}
\begin{equation}
\frac{f}{2}+f_{Q}\left[ 4\frac{\overset{.}{A}}{A}\frac{\overset{.}{B}}{B}%
+2\left( \frac{\overset{.}{B}}{B}\right) ^{2}\right] =\rho ,  \label{eqn18}
\end{equation}

\begin{equation}
\frac{f}{2}-f_{Q}\left[ -2\frac{\overset{.}{A}}{A}\frac{\overset{.}{B}}{B}-2%
\frac{\overset{..}{B}}{B}-2\left( \frac{\overset{.}{B}}{B}\right) ^{2}\right]
+2\frac{\overset{.}{B}}{B}\overset{.}{Q}f_{QQ}=-p,  \label{eqn19}
\end{equation}

\begin{equation}
\frac{f}{2}-f_{Q}\left[ -3\frac{\overset{.}{A}}{A}\frac{\overset{.}{B}}{B}-%
\frac{\overset{..}{A}}{A}-\frac{\overset{..}{B}}{B}-\left( \frac{\overset{.}{%
B}}{B}\right) ^{2}\right] +\left( \frac{\overset{.}{A}}{A}+\frac{\overset{.}{%
B}}{B}\right) \overset{.}{Q}f_{QQ}=-p,  \label{eqn20}
\end{equation}%
\end{widetext} where the dot $\left( \text{\textperiodcentered }\right) $
denotes the derivative with respect to cosmic time $\left( t\right) $. The
corresponding non-metricity scalar is given by \cite{ref27}

\begin{equation*}
Q=-2\left( \frac{\overset{.}{B}}{B}\right) ^{2}-4\frac{\overset{.}{A}}{A}%
\frac{\overset{.}{B}}{B}.
\end{equation*}

The field equations above (\ref{eqn18})-(\ref{eqn20}) can be represented in
the form of mean Hubble and directional Hubble parameters as

\begin{equation}
\frac{f}{2}-Qf_{Q}=\rho ,  \label{eqn21}
\end{equation}

\begin{equation}
\frac{f}{2}+2\frac{\partial }{\partial t}\left[ H_{y}f_{Q}\right]
+6Hf_{Q}H_{y}=-p,  \label{eqn22}
\end{equation}

\begin{equation}
\frac{f}{2}+\frac{\partial }{\partial t}\left[ f_{Q}\left(
H_{x}+H_{y}\right) \right] +3Hf_{Q}\left( H_{x}+H_{y}\right) =-p,
\label{eqn23}
\end{equation}%
where, we used $\frac{\partial }{\partial t}\left( \frac{\overset{.}{A}}{A}%
\right) =\frac{\overset{..}{A}}{A}-\left( \frac{\overset{.}{A}}{A}\right)
^{2}$ and $Q=-2H_{y}^{2}-4H_{x}H_{y}$. Here, $H=\frac{\overset{.}{a}}{a}=%
\frac{1}{3}\left( H_{x}+2H_{y}\right) $ is the Hubble parameter and $H_{x}=%
\frac{\overset{.}{A}}{A}$, $H_{y}=H_{z}=\frac{\overset{.}{B}}{B}$ represent
the directional Hubble parameters along $x$, $y$ and $z$ axes, respectively.

\section{Solutions for Bianchi type-I Universe}
\label{sec3}

In this section, we investigate solutions of the field equations for Bianchi
type-I space-time in $f\left( Q\right) $ gravity. Subtracting Eq. (\ref%
{eqn21}) from Eq. (\ref{eqn22}) we get,

\begin{equation}
\frac{d}{dt}\left[ f_{Q}\left( H_{x}-H_{y}\right) \right] +f_{Q}\left(
H_{x}-H_{y}\right) \frac{\overset{.}{V}}{V}=0,  \label{eqn24}
\end{equation}%
where $V=a^{3}=AB^{2}$ is the spatial volume of the Universe.\ By
integrating Eq. (\ref{eqn24}), we find

\begin{equation}
H_{x}-H_{y}=\frac{c_{1}}{Vf_{Q}},  \label{eqn25}
\end{equation}%
where $c_{1}$ is an integration constant. Afterward, we do some
manipulations, the field equations can be solved to get the metric
potentials of the form

\begin{equation}
A\left( t\right) =c_{2}^{\frac{2}{3}}V^{\frac{1}{3}}\exp \left[ \frac{2c_{1}%
}{3}\int \frac{dt}{Vf_{Q}}\right] ,  \label{eqn26}
\end{equation}

\begin{equation}
B\left( t\right) =c_{2}^{\frac{-1}{3}}V^{\frac{1}{3}}\exp \left[ \frac{-c_{1}%
}{3}\int \frac{dt}{Vf_{Q}}\right] .  \label{eqn27}
\end{equation}

The scalar expansion $\theta \left( t\right) $, shear scalar $\sigma
^{2}\left( t\right) $ and the mean anisotropy parameter $\Delta $ for
Bianchi type-I space-time are given by

\begin{equation}
\theta \left( t\right) =u_{;\mu }^{\mu }=3H,  \label{eqn28}
\end{equation}

\begin{equation}
\sigma ^{2}\left( t\right) =\frac{1}{2}\sigma _{ab}\sigma ^{ab}=\frac{1}{3}%
\left( H_{x}-H_{y}\right) ^{2},  \label{eqn29}
\end{equation}

\begin{equation}
\Delta =\frac{1}{3}\underset{i=1}{\overset{3}{\sum }}\left( \frac{H_{i}-H}{H}%
\right) ^{2}=6\left( \frac{\sigma }{\theta }\right) ^{2},  \label{eqn30}
\end{equation}%
where $H_{i}$ , $i=1,2,3$ are directional Hubble parameters.

Observations of the velocity redshift relation for extragalactic sources
indicate that the Hubble expansion of the Universe may reach isotropy when $%
\frac{\sigma }{\theta }$ is constant \cite{ref32}. Also, it has been
proposed that normal congruence to the homogeneous expansion for spatially
homogeneous metric produce $\frac{\sigma }{\theta }\approx 0.3$ \cite{ref33}%
. Bunn et al. \cite{ref34} performed a statistical analysis on the 4-year
CMB data and set a definitive upper bound on the amount of shear $\left( 
\frac{\sigma }{H}\right) $ for the primordial anisotropy to be less than $%
10^{-3}$ in the Planck era. As the Bianchi models constitute the anisotropic
space-time, one can choose that the ratio between the shear scalar and the
expansion scalar is constant $\left( \gamma \right) $ i.e. $\frac{\sigma }{%
\theta }=\gamma $. In this context, the condition was used in various
occasions in the papers \cite{ref35, ref36, ref37}.

From Eqs. (\ref{eqn25}) and (\ref{eqn30}), we have

\begin{equation}
\frac{\sigma }{\theta }=\frac{c_{1}}{\sqrt{3}\overset{.}{V}f_{Q}}=\gamma ,
\label{eqn31}
\end{equation}%
which implies that

\begin{equation}
f_{Q}=\frac{c_{1}}{\sqrt{3}\gamma \overset{.}{V}}.  \label{eqn32}
\end{equation}

To solve the field equations explicitly, we assume a volumetric expansion
law which is expressed as

\begin{equation}
V\left( t\right) =e^{3mt}t^{3n}.  \label{eqn33}
\end{equation}%
where $n\geq 0$\ and $m\geq 0$ are constants. This volumetric law gives the
scale factor as $a\left( t\right) =e^{mt}t^{n}$, which is called the hybrid
expansion law (HEL). It is known in cosmology as a generalization of the
power and exponential law. If $m=0$ this leads to the power law i.e. $%
a\left( t\right) =t^{n}$. As for $n=0$, this leads to the exponential law
i.e. $a\left( t\right) =e^{mt}$. This choice of scale factor produces a
time-dependent deceleration parameter, see \cite{ref38}.

For the hybrid expansion law with spatial volume given by Eq. (\ref{eqn33}),
one can get explicit relations of $A\left( t\right) $ and $B\left( t\right) $
in the following form

\begin{equation}
A\left( t\right) =c_{2}^{\frac{2}{3}}t^{n\left( 1+2\sqrt{3}\gamma \right)
}\exp \left[ m\left( 1+2\sqrt{3}\gamma \right) t\right] ,  \label{eqn34}
\end{equation}

\begin{equation}
B\left( t\right) =c_{2}^{\frac{-1}{3}}t^{n\left( 1-\sqrt{3}\gamma \right)
}\exp \left[ m\left( 1-\sqrt{3}\gamma \right) t\right] .  \label{eqn35}
\end{equation}

At the initial time, the metric potentials vanish. Thus, the model obtains
an initial singularity. As $t\rightarrow \infty $, the metric potentials
diverge to infinity. Consequently, there will be a big rip in the future,
because the $A\left( t\right) $ and $B\left( t\right) $ tends to infinity at 
$t\rightarrow \infty $.

Using Eqs. (\ref{eqn34}) and (\ref{eqn35}), we obtain $H$, $H_{x}$ and $%
H_{y} $ as

\begin{equation}
H\left( t\right) =m+\frac{n}{t},  \label{eqn36}
\end{equation}

\begin{equation}
H_{x}\left( t\right) =\left( 1+2\sqrt{3}\gamma \right) H\left( t\right) ,
\label{eqn37}
\end{equation}

\begin{equation}
H_{y}\left( t\right) =\left( 1-\sqrt{3}\gamma \right) H\left( t\right) .
\label{eqn38}
\end{equation}

The non-metricity scalar $Q$ in terms of the Hubble parameter for the hybrid
law model is found as follows

\begin{equation}
Q=-6\left( 1-3\gamma ^{2}\right) H^{2}\left( t\right) .  \label{eqn39}
\end{equation}

It is observed that the non-metricity of the Universe for our model is
time-dependent, because the Hubble parameter $H\left( t\right) $\ is a
function of cosmic time. As $t\rightarrow \infty $, the non-metricity scalar
tends to a constant value, i.e. $-6m^{2}\left( 1-3\gamma ^{2}\right) $.

The model of the cosmological constant or $\Lambda CDM$ model in GR is the
most successful model and the most widely used to date. Motivated by the
work of Solanki et al. \cite{ref39}, we assume the linear form of the $%
f\left( Q\right) $ function as

\begin{equation}
f\left( Q\right) =\alpha Q+\beta ,  \label{eqn40}
\end{equation}%
where $\alpha $ and $\beta $ are free model parameters.

Using the values of $Q$ and $H_{y}\left( t\right) $\ in Eqs. (\ref{eqn21})-(%
\ref{eqn23}), we obtain the energy density and pressure in terms of the
Hubble parameter as follows 
\begin{widetext}
\begin{equation}
\rho \left( t\right) =3\alpha \left( 1-3\gamma ^{2}\right) H^{2}\left(
t\right) +\frac{\beta }{2}.  \label{eqn41}
\end{equation}

\begin{equation}
p\left( t\right) =-3\alpha \left[ 1+3\gamma ^{2}\left( 1-\frac{2}{\sqrt{3}%
\gamma }\right) \right] H^{2}\left( t\right) -2\alpha \left( 1-\sqrt{3}%
\gamma \right) \overset{.}{H}\left( t\right) -\frac{\beta }{2},
\label{eqn42}
\end{equation}
\end{widetext} where, $\overset{.}{H}\left( t\right) =-\frac{n}{t^{2}}$. 
\newline

\begin{figure*}[tbp]
\begin{minipage}{0.45\linewidth}
  \centerline{\includegraphics[scale=0.7]{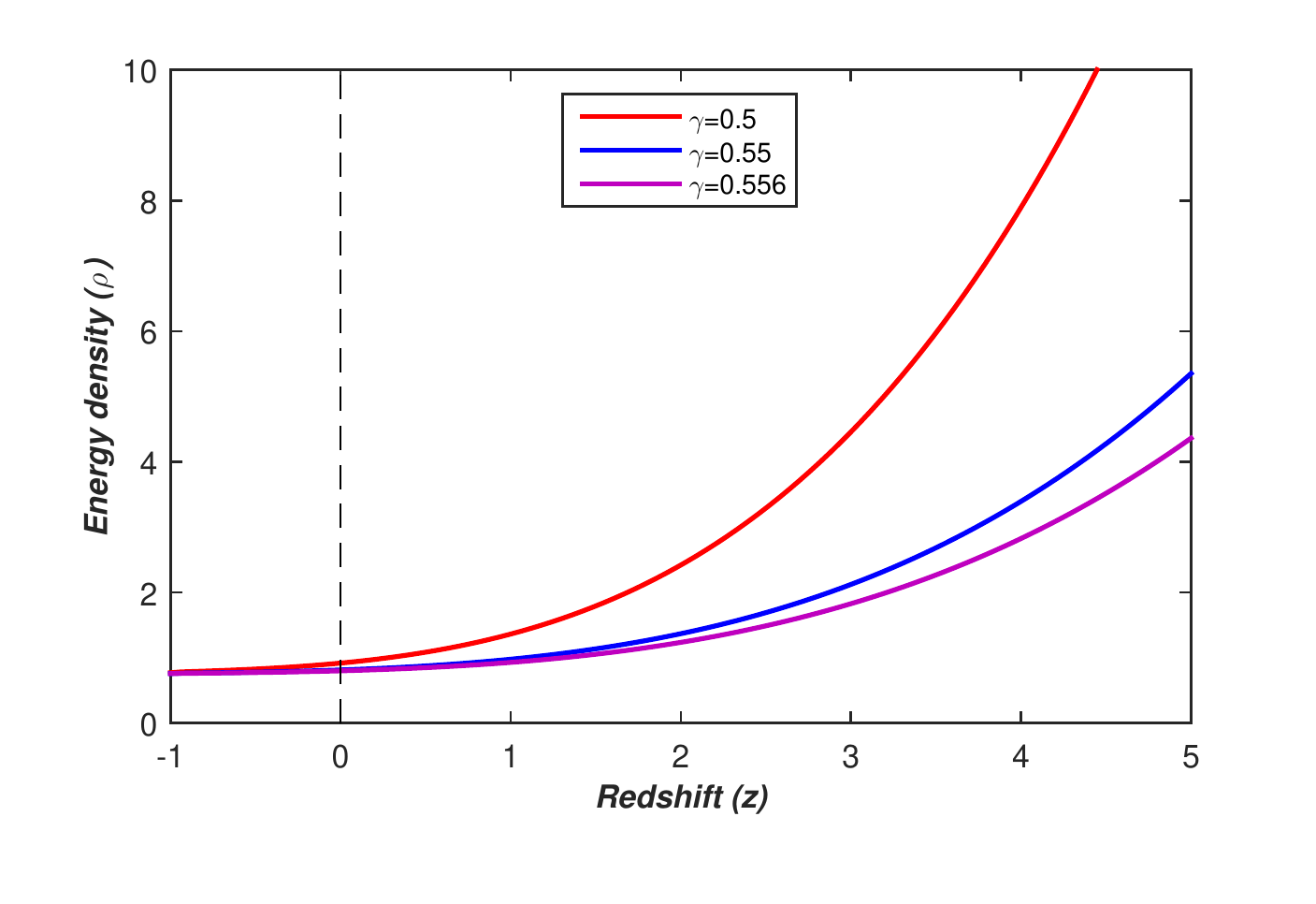}}
  \caption{{\emph{The energy density in terms of redshift for different values of $\gamma$.}}}\label{fig1}
 \end{minipage}\hfill 
\begin{minipage}{0.45\linewidth}
  \centerline{\includegraphics[scale=0.7]{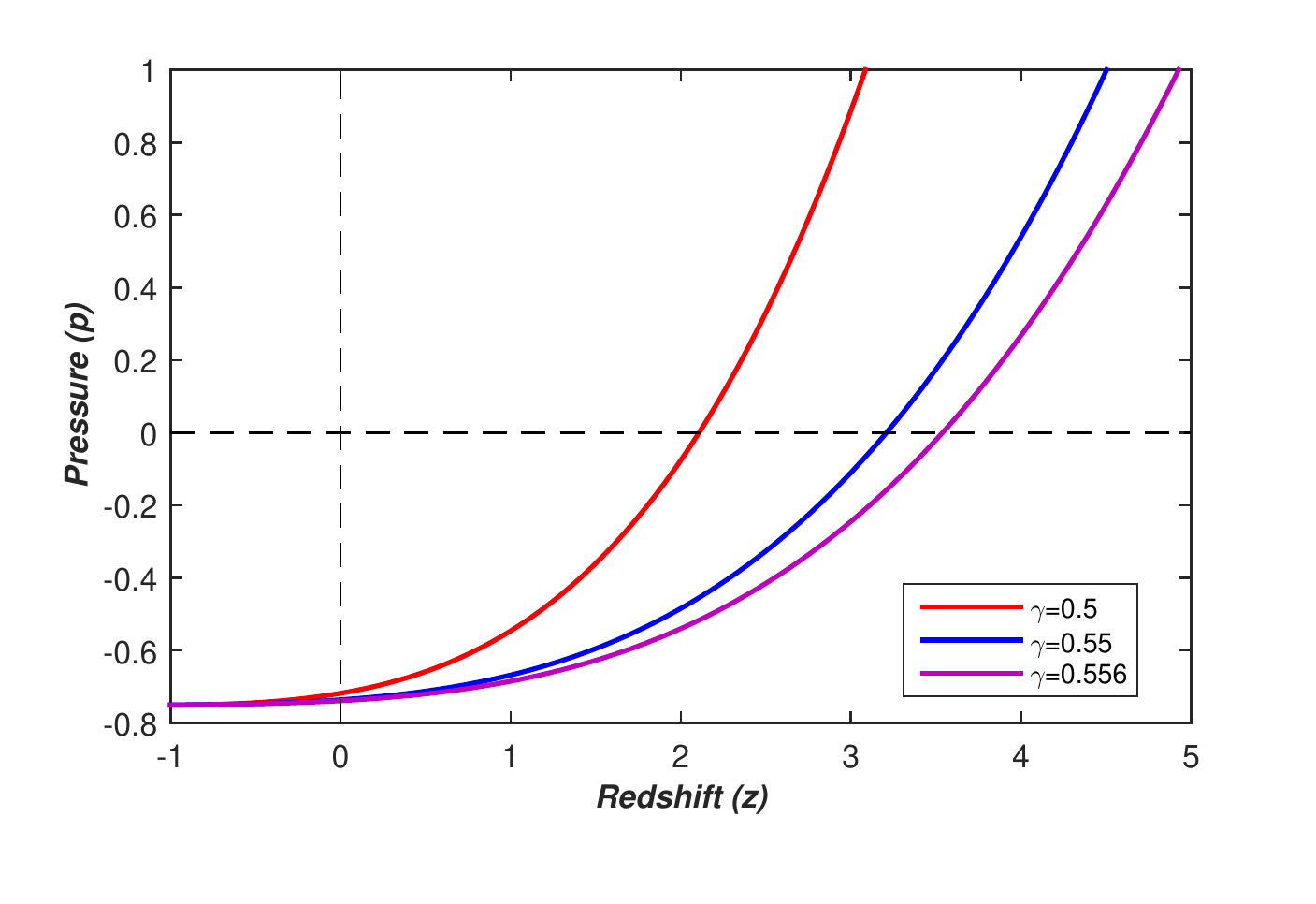}}
  \caption{{\emph{The pressure in terms of redshift for different values of $\gamma$.}}}\label{fig2}
 \end{minipage}
\end{figure*}

From above Eqs. (\ref{eqn41}) and (\ref{eqn42}), it is clear that $\rho
\left( t\right) $ and $p\left( t\right) $ $\rightarrow \infty $ as $%
t\rightarrow 0$. \ref{fig1} shows the behavior of the energy density of the
Universe in terms of redshift for different values of $\gamma $. We can see
that the variations of energy density versus redshift $z$ remains positive
throughout the evolution of the Universe and is an increasing function of $z$%
, or, equivalently, a decreasing function of cosmic time $t$. It is
initiated with a positive value and approaches to zero as $t\rightarrow
\infty $ (i.e. $z\rightarrow -1$). As for pressure, its behavior is shown in
Fig. \ref{fig2}, and is shown to be an increasing function of $z$, which starts from
a large positive value and tends to a negative value in the present and
future era. In reality, the negative pressure represents the accelerating
expansion of the present Universe. Thus, the Bianchi type-I space-time is
consistent with recent observational data of the Universe. Along with in
this analysis, we fix the said values of the model constants parameter
throughout the analysis: $m=0.5$, $n=0.6$, $\alpha =0.1$, and $\beta =1.5$.%
\newline

Using Eqs. (\ref{eqn41}) and (\ref{eqn42}), the EoS parameter of Bianchi
type-I space-time is obtained as 
\begin{widetext}
\begin{equation}
\omega \left( t\right) =\frac{p\left( t\right) }{\rho \left( t\right) }=-%
\frac{6\alpha \left[ 1+3\gamma ^{2}\left( 1-\frac{2}{\sqrt{3}\gamma }\right) %
\right] H^{2}\left( t\right) +4\alpha \left( 1-\sqrt{3}\gamma \right) 
\overset{.}{H}\left( t\right) +\beta }{6\alpha \left( 1-3\gamma ^{2}\right)
H^{2}\left( t\right) +\beta }.  \label{eqn43}
\end{equation}
\end{widetext}

\begin{table}[h]
\begin{center}
\begin{tabular}{c|c}
\hline
$\omega =0$ & dust \\ \hline
$\omega =\frac{1}{3}$ & radiation \\ \hline
$\omega \in \left( -\frac{1}{3},-1\right) $ & quintessence \\ \hline
$\omega =-1$ & cosmological constant \\ \hline
$\omega <-1$ & phantom matter \\ \hline
\end{tabular}%
\end{center}
\caption{{\emph{Various values of the EoS parameter.}}}
\label{tab1}
\end{table}

The EoS parameter is a very effective tool to study the evolutionary history
of the Universe, and we can summarize in the Tab. \ref{tab1} the different
values taken by the EoS parameter and its corresponding region. From Eq. (%
\ref{eqn43}), we have observed that the EoS parameter of Bianchi type-I
space-time is time-dependent. The graphical behavior of the EoS parameter
versus redshift $z$ is shown in Fig; 3. From this figure, we see that the
model starts in radiation dominated region, after which it passes from the
matter region and varies in the quintessence region, and finally, it
achieves $\Lambda CDM$ model for the values of $\gamma =0.5,0.55,0.556$. In
standard model $\left( \Lambda CDM\right) $, the Universe evolves from the
primordial phase dominated by photons, $\omega \sim \frac{1}{3}$, followed
by the period of the dominance of dust matter with $\omega =0$. Finally, the
cosmological constant dominates the Universe $\omega \rightarrow -1$ for
large values of time. Thus, the Bianchi type-I model corresponds to the $%
\Lambda CDM$ model. Also, the current values of the EoS parameter for the
three values of $\gamma $ are remarkably consistent with the most recent
Planck measurements (see Tab. \ref{tab2}) \cite{ref40}. 
\begin{figure}[tbp]
\centerline{\includegraphics[scale=0.7]{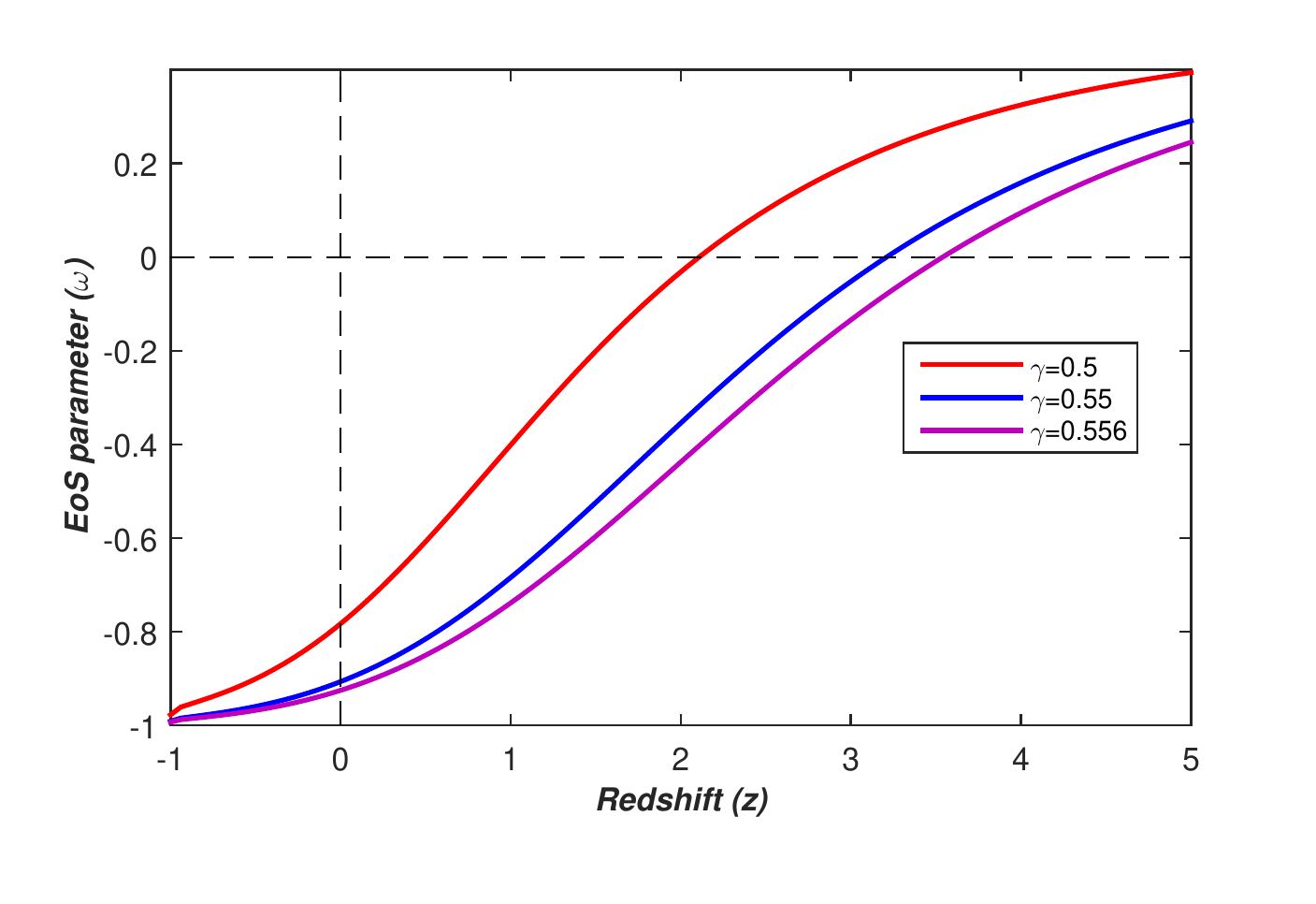}}
\caption{{\emph{The EoS parameter in terms of redshift for different values
of $\protect\gamma $.}}}
\label{fig3}
\end{figure}
\begin{table}[h]
\begin{tabular}{c|c}
\hline\hline
$\gamma $ & $\omega \left( z=0\right) $ \\ \hline\hline
$0.5$ & $-0.7827$ \\ \hline
$0.55$ & $-0.9065$ \\ \hline
$0.556$ & $-0.9255$ \\ \hline
\end{tabular}%
\caption{{\emph{The current values of the EoS parameter for the three values
of $\protect\gamma $.}}}
\label{tab2}
\end{table}

\section{Analysis of stability and energy conditions}
\label{sec4}
\textbf{Analysis of stability through the squared sound speed}:\newline

The squared sound speed $\left( v_{s}^{2}\right) $ is exploited for
examining the stability of the DE models which is given as $v_{s}^{2}=\frac{%
dp\left( t\right) }{d\rho \left( t\right) }$. If $v_{s}^{2}>0$, we obtain a
stable model and if $v_{s}^{2}<0$, we obtain unstable model. For Bianchi
type-I model,\ we observed the squared speed of sound, $v_{s}^{2}$ of the
form 
\begin{widetext}
\begin{equation}
v_{s}^{2}=-\frac{6\alpha \left[ 1+3\gamma ^{2}\left( 1-\frac{2}{\sqrt{3}%
\gamma }\right) \right] \overset{.}{H}\left( t\right) H\left( t\right)
+2\alpha \left( 1-\sqrt{3}\gamma \right) \overset{..}{H}\left( t\right) }{%
6\alpha \left( 1-3\gamma ^{2}\right) H\left( t\right) \overset{.}{H}\left(
t\right) }  \label{eqn44}
\end{equation}
\end{widetext} where $\overset{..}{H}\left( t\right) =\frac{2n}{t^{3}}$. 
\newline

\begin{figure}[tbp]
\centerline{\includegraphics[scale=0.7]{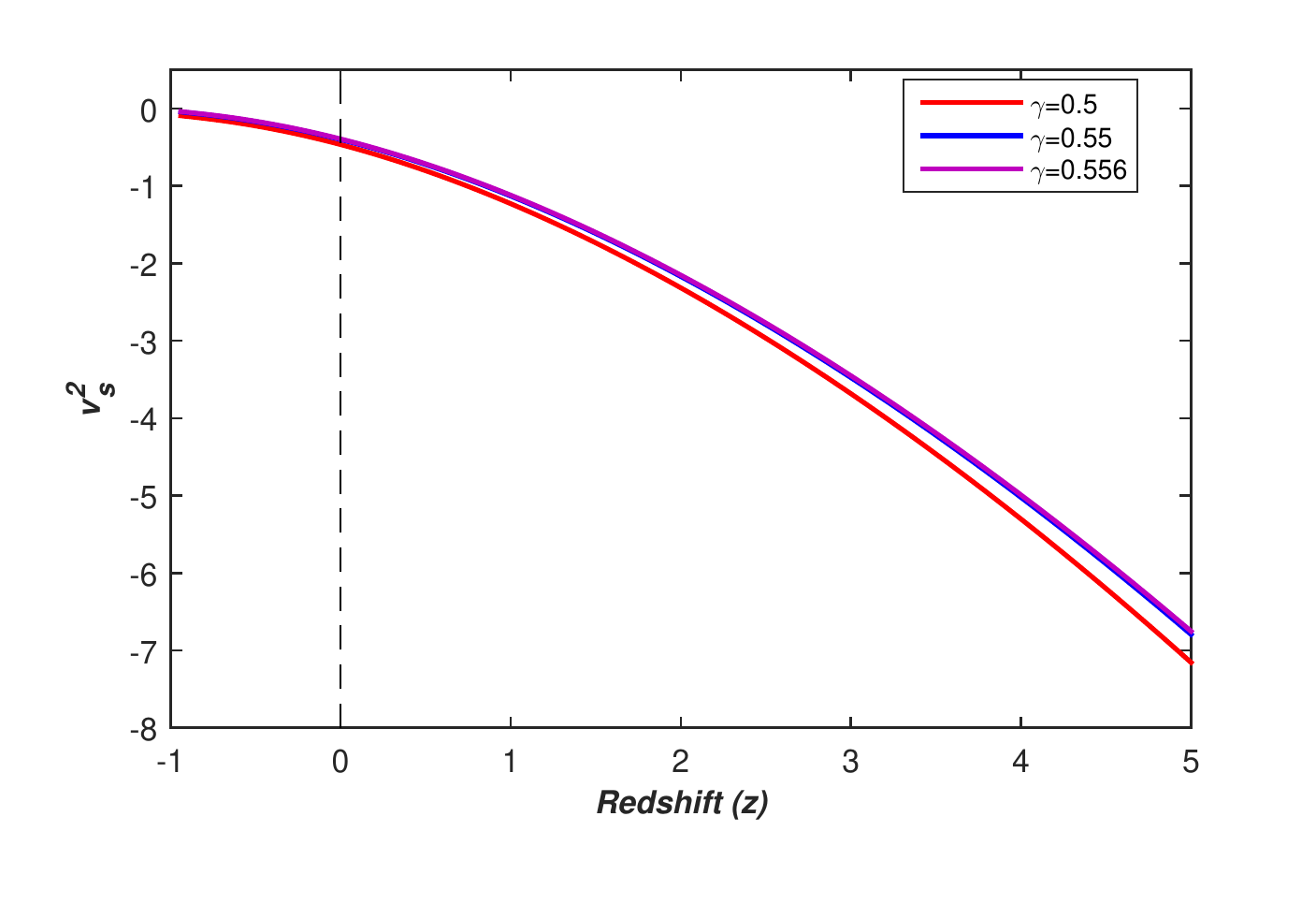}}
\caption{{\emph{The squared sound speed $\left( v_{s}^{2}\right) $ \ in
terms of redshift for different values of $\protect\gamma $.}}}
\label{fig4}
\end{figure}

Fig. \ref{fig4} shows the behavior of the squared speed of sound $v_{s}^{2}$ in terms
of redshift for different values of $\gamma $, and it can be noted from this
figure that the Bianchi type-I model is unstable throughout the cosmic
evolution.\newline

\textbf{Energy conditions}:\newline
\newline

The energy conditions are useful linear relationships composed of energy
density and pressure generated from the Raychaudhuri's equation. They are
important tools for validating dark energy models and are defined as follows 
\cite{ref20}:

\begin{itemize}
\item Weak energy conditions (WEC): $\rho \geq 0$,

\item Null energy condition (NEC): $\rho +p\geq 0$,

\item Dominant energy conditions (DEC): $\rho -p\geq 0$,

\item Strong energy conditions (SEC): $\rho +3p\geq 0$.
\end{itemize}

\begin{figure}[]
\centerline{\includegraphics[width=10cm,scale=1]{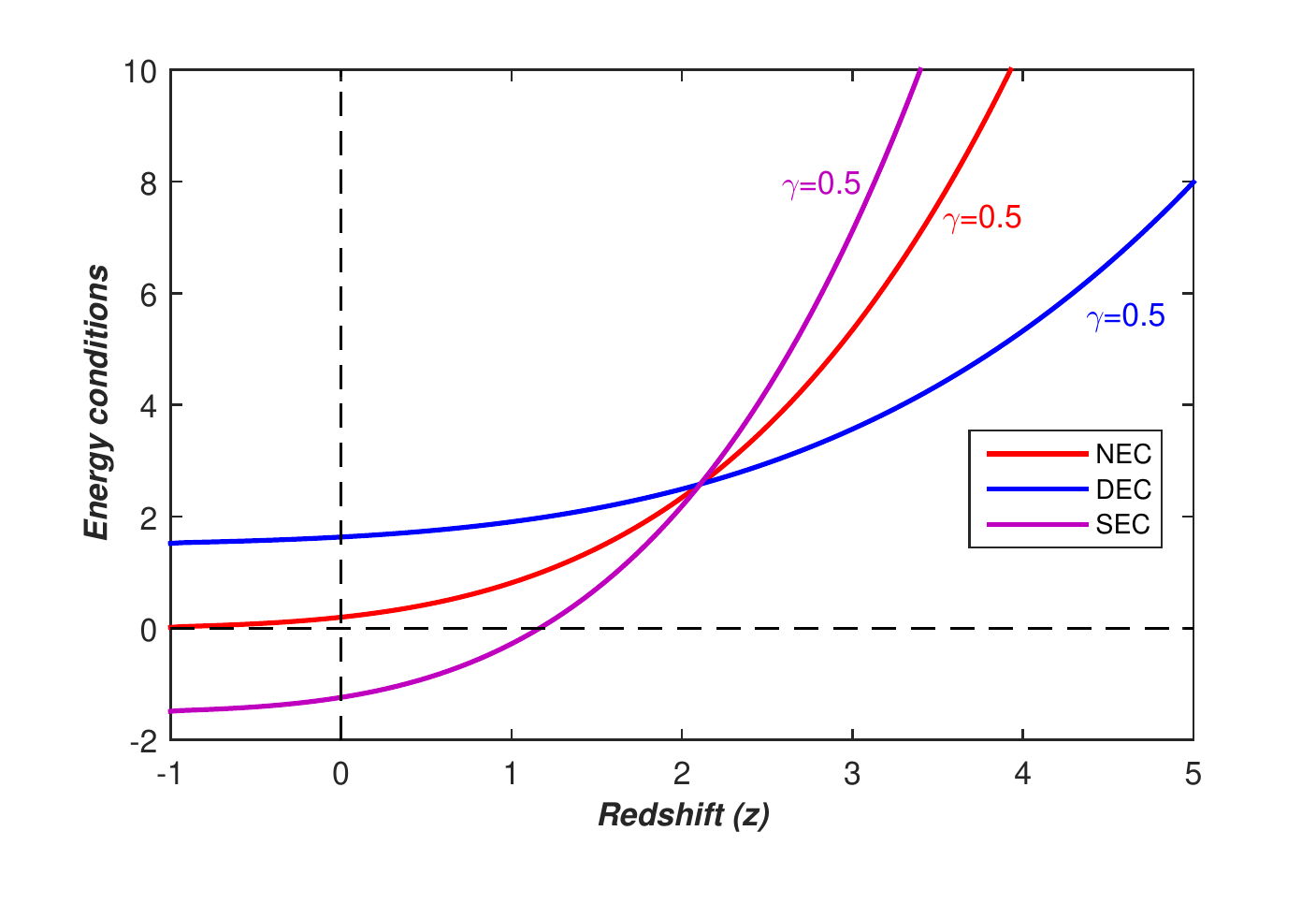}}
\caption{{\emph{The energy conditions in terms of redshift for $\protect%
\gamma=0.5$.}}}
\label{fig5}
\end{figure}

In Fig. \ref{fig5} we can see the evolution of energy conditions NEC, DEC, and SEC in
terms of redshift. In order to explicate the late-time cosmic acceleration
with $\omega \simeq -1$, the SEC is necessary to violate i.e. $\rho \left(
1+3\omega \right) \leq 0$. It is clearly shown by Fig. \ref{fig5} that this condition
is being violated in the present era and in the future.

\section{Cosmological parameters in terms of redshift}
\label{sec5}

In the previous sections we have plotted the behavior of all physical
parameters in terms of redshift $z$, in fact we used here the relationship
between the scale factor and the redshift, which is expressed by

\begin{equation}
a\left( t\right) =\frac{a\left( t_{0}\right) }{1+z},  \label{eqn45}
\end{equation}%
where $a\left( t_{0}\right) $ is the present scale factor. \newline

For our model, we find the time-redshift relation in the form \cite{ref38}

\begin{equation}
t\left( z\right) =\frac{nW\left[ \frac{m\left( \frac{1}{1+z}\right) ^{\frac{1%
}{n}}}{n}\right] }{m},  \label{eqn46}
\end{equation}%
where $W$ denotes the Lambert function (also known as \textquotedblleft
product logarithm\textquotedblright ).

The deceleration parameter is the quantity that describes the evolution of
the expansion of the Universe. This parameter is positive $\left( q>0\right) 
$ when the Universe decelerates over time, and is negative $\left(
q<0\right) $ in an accelerated expanding Universe. This parameter is related
to the Hubble parameter as follows $q=-1-\frac{\overset{.}{H}}{H^{2}}$.
Thus, the Hubble parameter and deceleration parameter of our model in terms
of redshift are respectively obtained as

\begin{equation}
H\left( z\right) =\frac{m}{W\left[ \frac{m\left( \frac{1}{1+z}\right) ^{%
\frac{1}{n}}}{n}\right] }+m.  \label{eqn47}
\end{equation}

\begin{equation}
q\left( z\right) =-1+\frac{1}{n\left( 1+W\left[ \frac{m\left( \frac{1}{1+z}%
\right) ^{\frac{1}{n}}}{n}\right] \right) ^{2}}.  \label{eqn48}
\end{equation}

From (\ref{eqn48}), we observe that

\begin{center}
\begin{eqnarray}
q &<&0\ \ if\ \ t>\frac{\sqrt{m}-m}{n},  \label{eqn49} \\
q &>&0\ \ if\ \ t<\frac{\sqrt{m}-m}{n}.  \notag
\end{eqnarray}
\end{center}

\begin{figure*}[tbp]
\begin{minipage}{0.45\linewidth}
  \centerline{\includegraphics[scale=0.7]{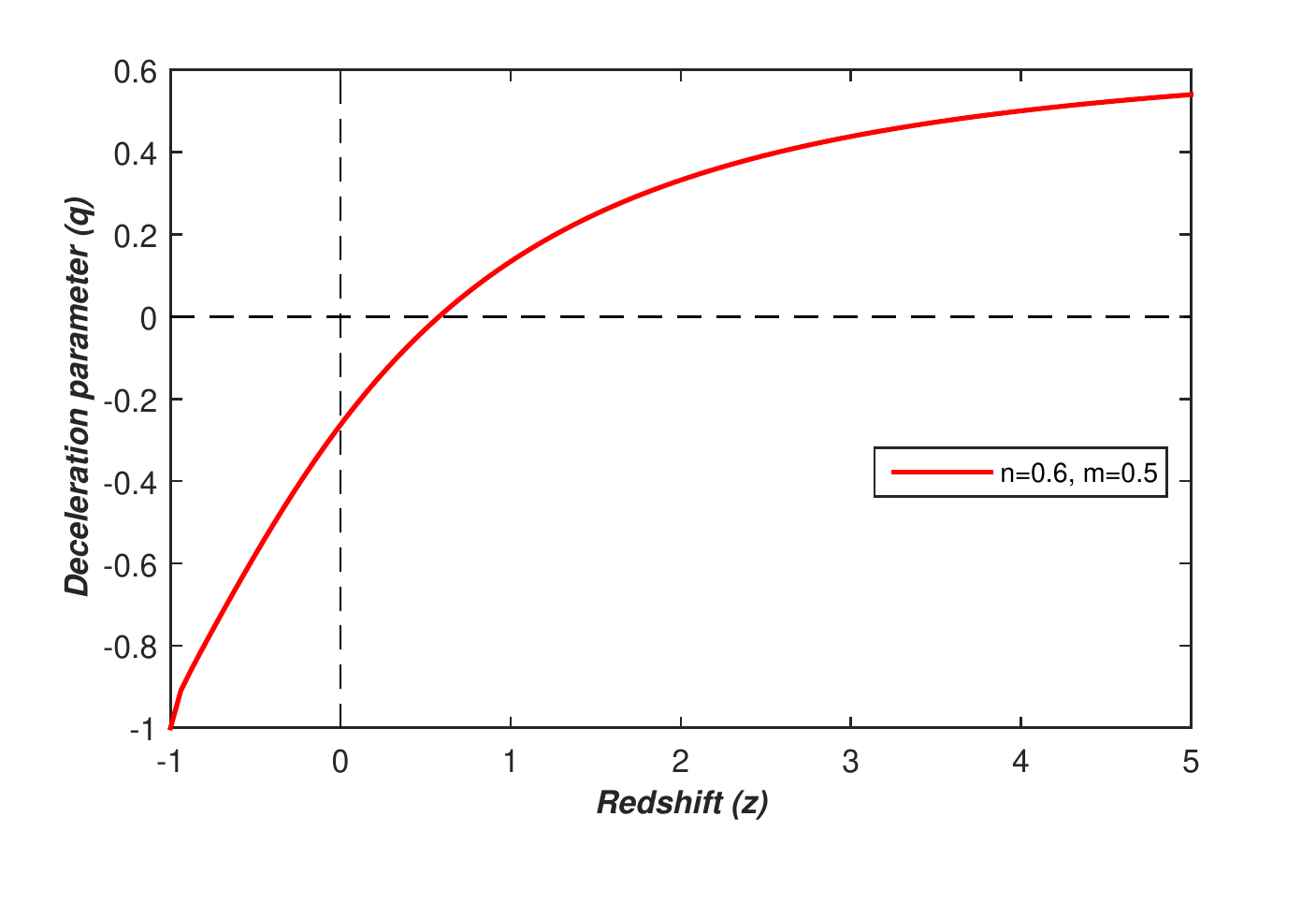}}
 \caption{{\emph{The deceleration parameter in terms of redshift.}}}\label{fig6}
 \end{minipage}\hfill 
\begin{minipage}{0.45\linewidth}
  \centerline{\includegraphics[scale=0.7]{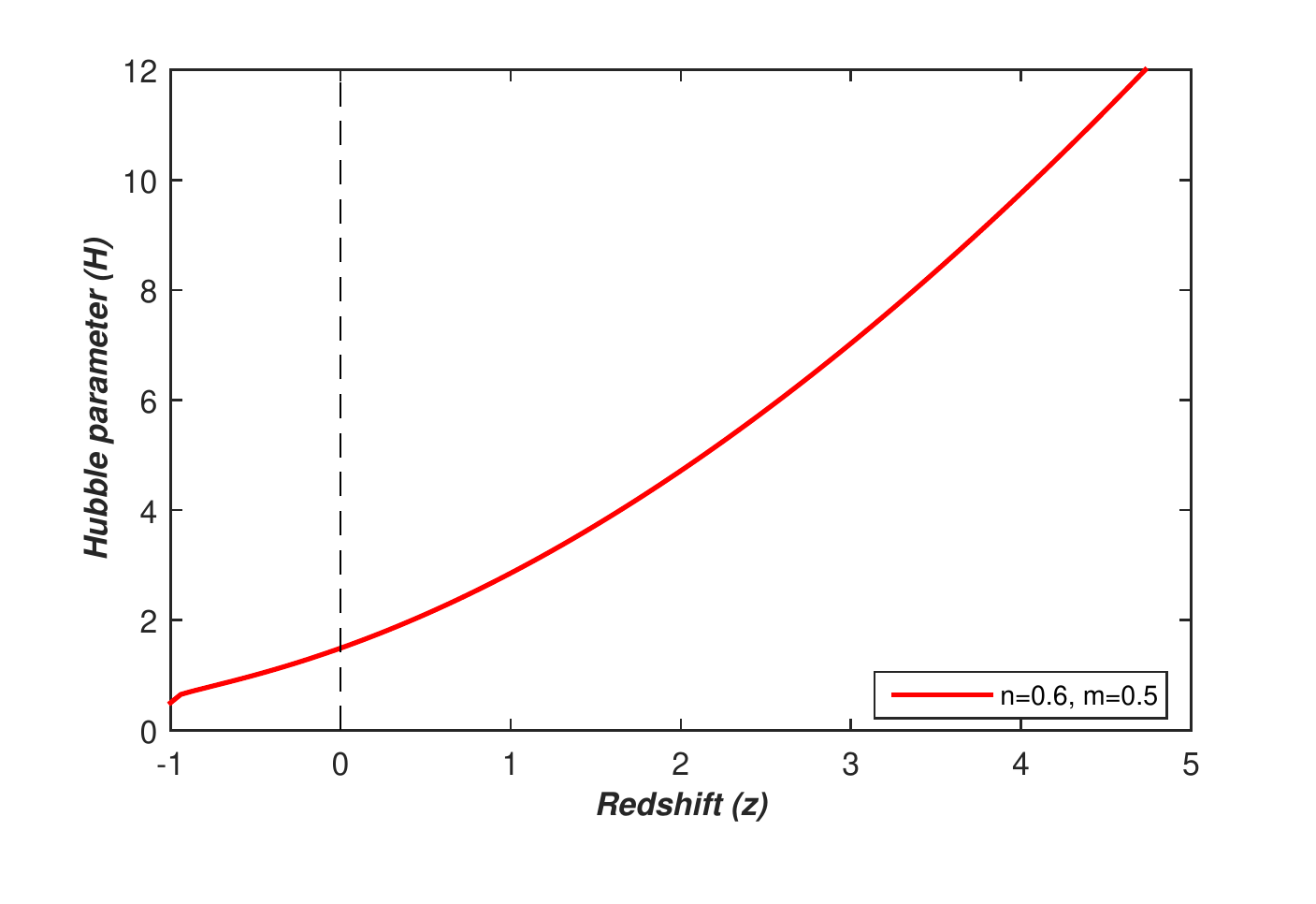}}
 \caption{{\emph{The Hubble parameter in terms of redshift.}}}\label{fig7}
 \end{minipage}
\end{figure*}

Recent observations of SN-Ia revealed that the Universe in the current era
is accelerating and that the deceleration parameter value is in the range $%
-1\leq q<0$. Figs. 6 and 7 show the variation of Hubble parameter $H\left(
z\right) $ and deceleration parameter $q\left( z\right) $ versus redshift $z$%
. It can be seen from Fig. \ref{fig} that $q\left( z\right) $ decreases from
positive to negative zone and finally tends to $-1$. Thus, our model shows a
transition from decelerated phase to an accelerated phase. Also, the value
of the deceleration parameter is consistent with the observational data.

\section{best fit values of model parameters from observation ($H(z)$
datasets)}

As the cosmological principle on the large scale, universe is homogeneous
and isotropic be the backbone of modern cosmology which had been tested
several times by the researchers and supported by many cosmological
observations. In this study, the expansion scenario of the universe be
directly investigated by the Hubble parameter as a function of redshift. To
measure the value of the Hubble parameter at some definite redshift
generally two well-known methods such as \textit{the extraction of H(z) from
line-of-sight BAO data} and \textit{the differential age method} are used 
\cite{ref41}. In order to find the best fit value of the model parameters of
our obtained model, we have used the technique of $57$ points of Hubble
parameter values $H(z)$ with $\sigma _{\mu }^{2}$ errors of differential age
method ($31$ points) and $BAO$ and other methods ($26$ points). Through the
scale factor proposed in this model given in Eq. (\ref{eqn33}), we find its
expression at the present time as $a_{0}=e^{mt_{0}}t_{0}^{n}$ where $t_{0}$
represent the present age of the Universe. Thus, after a simple calculation,
we find the scale factor for our model as follows: $a(t)=a_{0}e^{k\left( 
\frac{t}{t_{0}}-1\right) }\left( \frac{t}{t_{0}}\right) ^{n}$ where $%
k=mt_{0} $. The Hubble parameter in terms of redshift of our model with
recent modification reads as 
\begin{equation}
H(z)=\frac{H_{0}k}{k+n}\left[ W^{-1}\left( \frac{k}{n}e^{\frac{k-ln(1+z)}{n}%
}\right) +1\right]  \label{eqn50}
\end{equation}%
where $H_{0}=\frac{k+n}{t_{0}}$. We see from the above equation that the
model parameters consist of three parameters $k,n$ and $H_{0}$ which will be
constrained hereinafter by the last observations of Hubble datasets. The
corresponding $\chi $-square function to obtain the best adjustment value
for the parameters $k,n$ and $H_{0}$ is determined by 
\begin{equation}
\chi _{OHD}^{2}(k,n,H_{0})=\sum_{\mu =1}^{57}\frac{\left[
H_{th}(k,n,H_{0},z_{\mu })-H_{obs}(z_{\mu })\right] ^{2}}{\sigma _{\mu }^{2}}
\label{eqn51}
\end{equation}%
where $H_{th}(k,n,H_{0},z_{\mu })$ and $H_{obs}(z_{\mu })$ are theoretical
and observed values of Hubble parameter respectively, and $\sigma _{\mu
}^{2} $ represents the standard error in the observed Hubble parameter
measurements. $\sigma _{\mu }^{2}$ errors of differential age method ($31$
points) and $BAO$ and other methods ($26$ points) are represented in Tab.
III.\newline
\begin{table}[tbp]
\begin{center}
\begin{tabular}{|c|c|c|c|c|c|}
\hline\hline
$z$ & $H(z)$ & $\sigma _{H}$ & $z$ & $H\left( z\right) $ & $\sigma _{H}$ \\ 
\hline\hline
0.070 & 69 & 19.6 & 0.4783 & 80 & 99 \\ \hline
0.90 & 69 & 12 & 0.480 & 97 & 62 \\ \hline
0.120 & 68.6 & 26.2 & 0.593 & 104 & 13 \\ \hline
0.170 & 83 & 8 & 0.6797 & 92 & 8 \\ \hline
0.1791 & 75 & 4 & 0.7812 & 105 & 12 \\ \hline
0.1993 & 75 & 5 & 0.8754 & 125 & 17 \\ \hline
0.200 & 72.9 & 29.6 & 0.880 & 90 & 40 \\ \hline
0.270 & 77 & 14 & 0.900 & 117 & 23 \\ \hline
0.280 & 88.8 & 36.6 & 1.037 & 154 & 20 \\ \hline
0.3519 & 83 & 14 & 1.300 & 168 & 17 \\ \hline
0.3802 & 83 & 13.5 & 1.363 & 160 & 33.6 \\ \hline
0.400 & 95 & 17 & 1.430 & 177 & 18 \\ \hline
0.4004 & 77 & 10.2 & 1.530 & 140 & 14 \\ \hline
0.4247 & 87.1 & 11.2 & 1.750 & 202 & 40 \\ \hline
0.4497 & 92.8 & 12.9 & 1.965 & 186.5 & 50.4 \\ \hline
0.470 & 89 & 34 &  &  &  \\ \hline\hline\hline
$z$ & $H\left( z\right) $ & $\sigma _{H}$ & $z$ & $H\left( z\right) $ & $%
\sigma _{H}$ \\ \hline\hline
0.24 & 79.69 & 2.99 & 0.52 & 94.35 & 2.64 \\ \hline
0.30 & 81.7 & 6.22 & 0.56 & 93.34 & 2.3 \\ \hline
0.31 & 78.18 & 4.74 & 0.57 & 87.6 & 7.8 \\ \hline
0.34 & 83.8 & 3.66 & 0.57 & 96.8 & 3.4 \\ \hline
0.35 & 82.7 & 9.1 & 0.59 & 98.48 & 3.18 \\ \hline
0.36 & 79.94 & 3.38 & 0.60 & 87.9 & 6.1 \\ \hline
0.38 & 81.5 & 1.9 & 0.61 & 97.3 & 2.1 \\ \hline
0.40 & 82.04 & 2.03 & 0.64 & 98.82 & 2.98 \\ \hline
0.43 & 86.45 & 3.97 & 0.73 & 97.3 & 7.0 \\ \hline
0.44 & 82.6 & 7.8 & 2.30 & 224 & 8.6 \\ \hline
0.44 & 84.81 & 1.83 & 2.33 & 224 & 8 \\ \hline
0.48 & 87.90 & 2.03 & 2.34 & 222 & 8.5 \\ \hline
0.51 & 90.4 & 1.9 & 2.36 & 226 & 9.3 \\ \hline
\end{tabular}%
\end{center}
\caption{57 points of $H(z)$ data: $31$ (DA) and $26$ (BAO+other) 
\protect\cite{ref41}.}
\end{table}

Now, using $57$ Hubble parameter measurements we will try to find the best
fit values for the parameters model $k,n$ and $H_{0}$ where the best fit
curve for the Hubble parameter versus redshift $z$ is represented in Fig. \ref{fig8}.
Thus, The best fit values are gained as $k=0.2239\pm 0.3201$, $n=0.6886\pm
0.2868$ and $H_{0}=62.73\pm 8.43$ $km.s^{-1}.Mpc^{-1}.$

\begin{figure}[tbp]
\centerline{\includegraphics[scale=0.4]{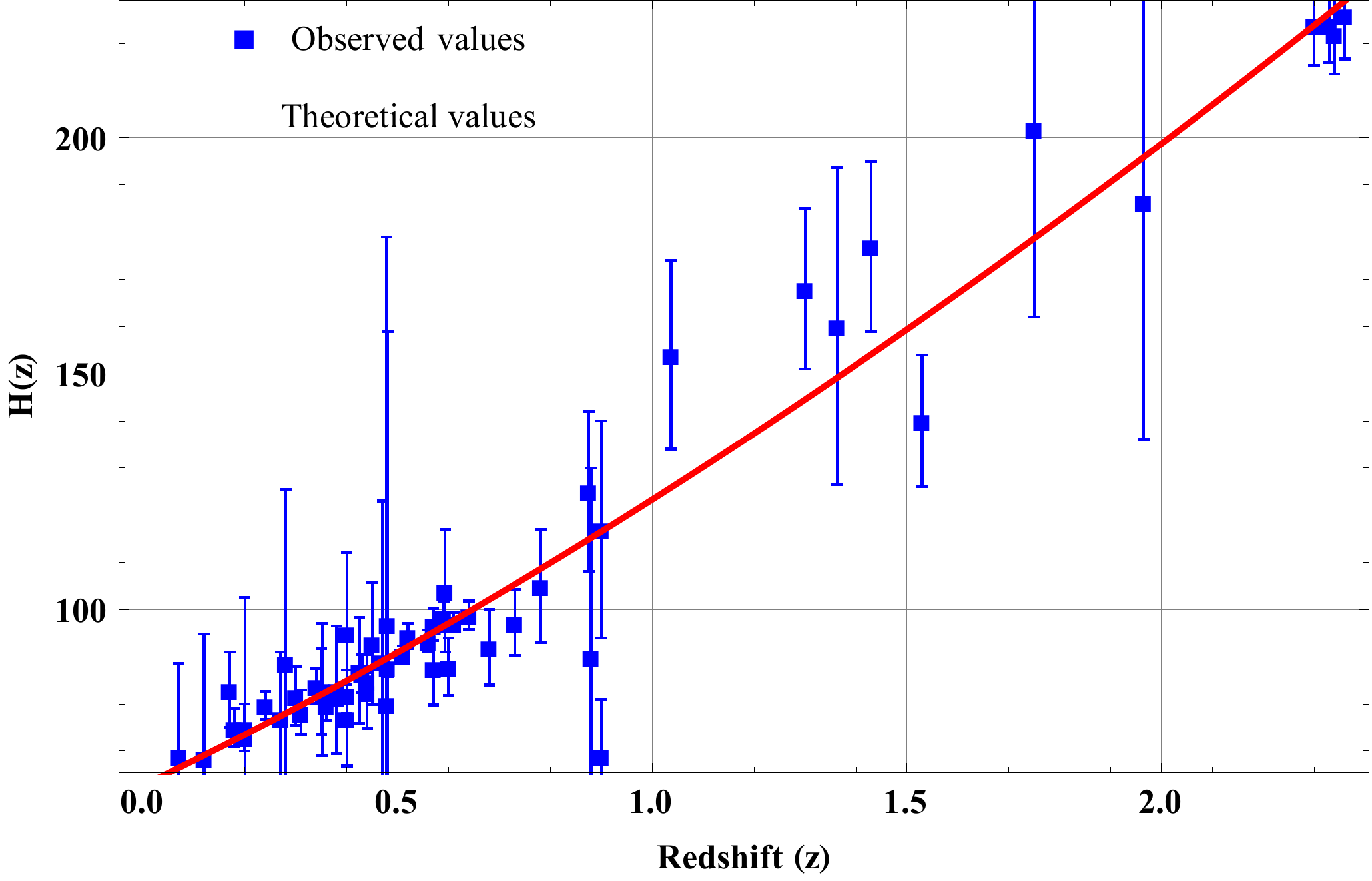}}
\caption{{\emph{Best fit curve of Hubble function $H(z)$ vs. redshift $z$.}}}
\end{figure}

\section{Conclusion and discussion}
\label{sec6}

In this analysis, we have investigated an anisotropic Bianchi type-I
space-time with perfect fluid in the framework of $f\left( Q\right) $
gravity, where $Q$ is the non-metricity scalar. Motivated by the work of
Solanki et al. \cite{ref31} we choose the $f\left( Q\right) $\ function on
the linear form, i.e. $f\left( Q\right) =\alpha Q+\beta $, where $\alpha $
and $\beta $ are free model parameters. We have used the ratio between the
shear scalar and the expansion scalar is constant i.e. $\frac{\sigma }{%
\theta }=\gamma $. Here, we have examined the behavior of our model for
three different values of $\gamma =0.5,0.55,0.556$.

It is observed that the metric potentials $A\left( t\right) $ and $B\left(
t\right) $ vanish at the initial time (at $t=0$), and diverge at $%
t\rightarrow \infty $. Thus, the model obtains the initial singularity and
predicts a big rip in the future. From Fig. \ref{fig1}, we see that the energy
density of the Bianchi type-I Universe is a positive and decreasing function
of cosmic time. Also, $\rho \rightarrow const$ as $t\rightarrow \infty $,
which leads the fact that the volume of the space increases because of the
density of matter decreases as the Universe expands in our model. The
pressure in Fig. \ref{fig2} takes negative values in the present and future eras, and
this produces the current phase of the acceleration of the Universe, which
is governed by dark energy. In Fig. \ref{fig3}, we can see that the Universe is
dominated by radiation $\left( \omega \sim \frac{1}{3}\right) $ in the first
epoch, and later it passes from the matter-dominated dust epoch $\left(
\omega =0\right) $ to the Quintessence region $\omega \in \left( -\frac{1}{3}%
,-1\right) $ and finally reaches the cosmological constant $\Lambda $ phase
which causes the acceleration cosmic i.e. $\omega =-1$. We conclude that the
model behaves like the standard cosmological model. To compare the current
values of the EoS parameter obtained with this model with the recent Hubble
measurements, we mention that Aghanim et al. \cite{ref39} found constraint
for the EoS parameter as follows:

\begin{itemize}
\item $\omega =-1.56_{-0.84}^{+0.60}$ (Planck + TT + lowE),

\item $\omega =-1.58_{-0.41}^{+0.52}$ (Planck + TT, EE + lowE),

\item $\omega =-1.57_{-0.40}^{+0.50}$ (Planck + TT, TE, EE + lowE + lensing),

\item $\omega =-1.04_{-0.10}^{+0.10}$ (Planck + TT, TE, EE + lowE + lensing
+ BAO).
\end{itemize}

In Tab. II we provided the observed values in the derived model which are in
excellent agreement with the above values.

In Section 4 we have examined the behavior of energy conditions in terms of
the redshift of a Bianchi type-I space-time and found that WEC and NEC are
all satisfied but SEC is violated. This leads directly to the creation of an
accelerating phase of the Universe because $\rho \left( 1+3\omega \right) <0$
results in that $\omega \simeq -1$, which corresponds to the observations.
Further, for constraining the model parameter we have used Hubble data data
sets ($57$ points of Hubble parameter values $H(z)$ see Tab. III). Hence,
from the Hubble datasets, we have the best fit ranges for the model
parameters are $k=0.2239\pm 0.3201$, $n=0.6886\pm 0.2868$ and $%
H_{0}=62.73\pm 8.43$.\newline
Also, we have discussed the analysis of stability using the squared sound
speed $\left( v_{s}^{2}\right) $ and found that the model is unstable
throughout cosmic evolution i.e. $v_{s}^{2}<0$. In Figs.s 6 and 7, we draw
the evolution of the deceleration parameter and the Hubble parameter in
terms of the redshift, and it appears that the model passes from the
deceleration phase to the current accelerated phase, and in the latter, it
yields to $-1$. This result is consistent with the observational data of
recent Planck measurements.

\section*{Acknowledgments}

We the authors are very much grateful to the editor and anonymous referee
for illuminating suggestions that have significantly improved the quality
the article. Also, would like to thank our colleagues from EL-AKKAD HIGH
SCHOOL for their moral support and especially the English teacher, Mr. S.
Elkaoukaji who helped us to write this article.

\textbf{Data availability} There are no new data associated with this article%

\textbf{Declaration of competing interest} The authors declare that they
have no known competing financial interests or personal relationships that
could have appeared to influence the work reported in this paper.\newline

\bigskip

\end{document}